\documentclass[%
 aip,
 amsmath,amssymb,
 reprint,%
]{revtex4-1}

\usepackage{graphicx}
\usepackage{dcolumn}
\usepackage{bm}
\usepackage{float} 

\usepackage[utf8]{inputenc}
\usepackage[T1]{fontenc}
\usepackage{mathptmx}
\usepackage{etoolbox}
\usepackage{color,soul} 
\usepackage[colorlinks=true, linkcolor=blue, citecolor=blue, urlcolor=blue]{hyperref}
\usepackage{microtype}
\makeatletter
\def\@email#1#2{%
 \endgroup
 \patchcmd{\titleblock@produce}
  {\frontmatter@RRAPformat}
  {\frontmatter@RRAPformat{\produce@RRAP{*#1\href{mailto:#2}{#2}}}\frontmatter@RRAPformat}
  {}{}
}%
\makeatother
\begin{document}

\preprint{AIP/123-QED}

\title[]{Efficient detection of spectrally multimode squeezed light \\ through optical parametric amplification}
\author{M. Kalash}
 
\affiliation{ 
Friedrich-Alexander Universit\"at Erlangen-N\"urnberg, Staudtstra\ss{}e 7/B2, 91058 Erlangen, Germany
}
\affiliation{Max Planck Institute for the Science of Light, Staudtstra\ss{}e 2, 91058 Erlangen, Germany}
\author{U. Han}%
\affiliation{%
Department of Physics, Korea Advanced Institute of Science and Technology (KAIST), Daejeon 34141, Korea}%
\author{Y.-S. Ra}
\affiliation{%
Department of Physics, Korea Advanced Institute of Science and Technology (KAIST), Daejeon 34141, Korea}%
\author{M. V. Chekhova}

\affiliation{ 
Friedrich-Alexander Universit\"at Erlangen-N\"urnberg, Staudtstra\ss{}e 7/B2, 91058 Erlangen, Germany
}%
\affiliation{Max Planck Institute for the Science of Light, Staudtstra\ss{}e 2, 91058 Erlangen, Germany}
 \email{mahmoud.kalash@mpl.mpg.de}

\date{\today}

\begin{abstract}
Multimode squeezed light is a key resource for high-dimensional photonic quantum technologies, enabling applications in quantum-enhanced sensing, quantum communication, and quantum computing. Efficient detection of such a multimode squeezed state is essential for unlocking its full potential. Optical parametric amplification (OPA) has recently gained attention as a powerful technique offering loss-tolerant, direct broadband detection, and multimode operation. While OPA has been used to characterize spatially multimode squeezing, its application to spectrally multimode squeezing has not yet been demonstrated. Here, we report on the first experimental demonstration of spectrally multimode squeezing detection using OPA. We achieve simultaneous detection of squeezing across more than 60 spectral modes of a broadband squeezed vacuum state. The observed squeezing is nearly uniform, ranging from -6.5 to -7 dB, which makes the source particularly suitable for constructing continuous-variable cluster states, and indicates the multimode capability of the OPA. The results extend the capabilities of OPA detection into the spectral domain, advancing spectral-mode-based high-dimensional photonic quantum technologies.
\end{abstract}

\maketitle

\section{\label{intro}Introduction}
Optical parametric amplification (OPA) has recently gained attention as a powerful tool for the detection of quantum states of light. It has been proposed for ultrafast optical quantum processors\cite{inoueMulticoreUltrafastOptical2023}, for single-photon detection~\cite{Levenson1993,Sendonaris2024}, and applied to the certification of non-Gaussianity and non-classicality of quantum states without the need for single-photon detectors~\cite{Kalash2025}. In addition, OPA has been implemented for full quantum state tomography~\cite{Kalash2023Sep}. Specifically for squeezed light, by amplifying one quadrature of the optical field while de-amplifying its conjugate, OPA enables direct mapping of quadrature variances to output intensity, offering loss-tolerant, all-optical, and broadband detection of squeezing~\cite{Shaked2018Feb,Frascella2019Sep,Takanashi2020Nov,Nehra2022Sep}. These advances highlight OPA’s emerging role as a versatile tool across a wide range of optical quantum applications.

Crucially, OPA offers a unique advantage due to its inherent multimode nature. A multimode OPA (MOPA) is capable of simultaneously amplifying multiple spatial and spectral modes, making it well-suited for detecting multimode squeezed states where conventional homodyne detection (HD) faces challenges due to to mode-selective measurements and sensitivity to loss. Recent works have experimentally demonstrated OPA-based detection of spatially multimode squeezing, including post-processed measurements~\cite{Barakat2025Feb} as well as real-time monitoring~\cite{Kalash2025Mar} of squeezing in multiple co-propagating spatial modes. These successes have encouraged the integration of MOPA detection into the spectral domain, where squeezed light across frequency modes plays a key role in different applications\cite{chang2025,kues2019,roh2025}. 

In this work, we present the first experimental demonstration of spectrally multimode squeezing detection using MOPA. We achieve simultaneous detection of squeezing across 60 spectral modes of a broadband squeezed vacuum state. We achieved uniform squeezing ranging between –7 dB and -6.5 dB with high purity. This demonstration extends the capabilities of OPA-based detection into the spectral domain, significantly broadening its potential for quantum technologies, particularly in platforms leveraging optical frequency combs~\cite{herman2025,cai2017,yang2021,zhao2020,guidry2023,jahanbozorgi2023,roh2025}.
\section{\label{MOPA}Detection of multimode squeezing through MOPA}
\begin{figure}[h!]
\centering
\includegraphics[width=0.7\linewidth,height=0.5\linewidth]{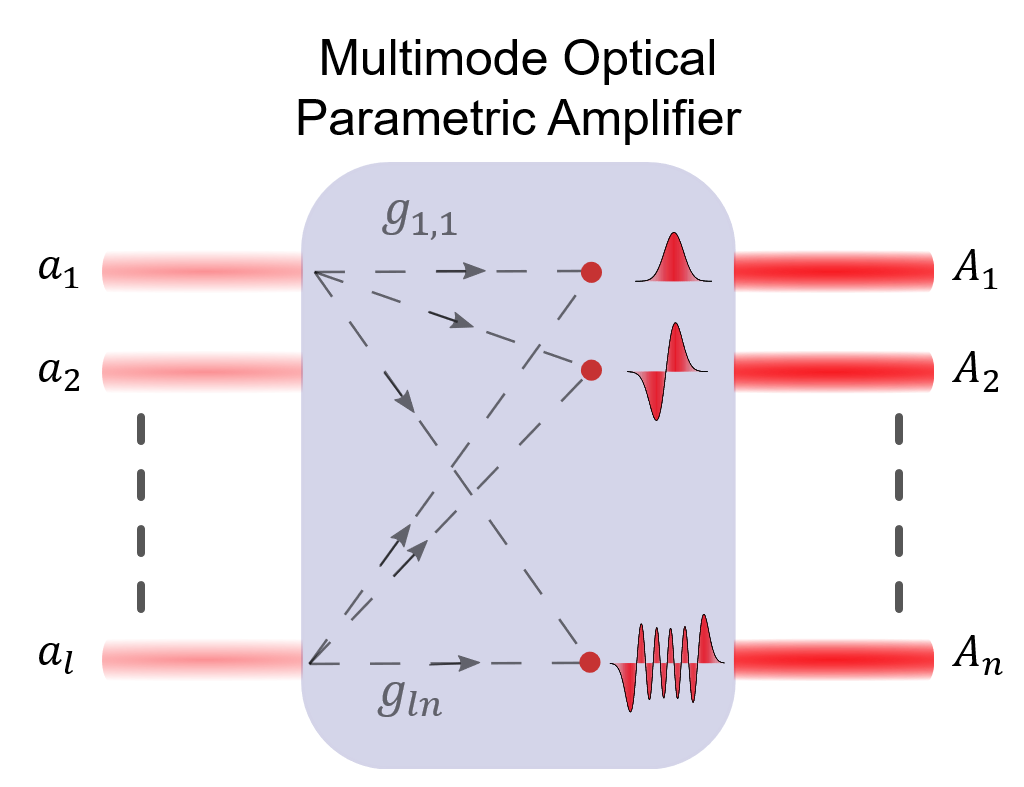}
     \caption{Multimode optical parametric amplification with $a_l$ and $A_n$ are fields occupying input and output modes, respectively.  $g_{ln}$ denotes the input-output mode overlap.} 
    \label{idea}
\end{figure}
\begin{figure*}[t!]
\centering
\includegraphics[width=0.7\linewidth]{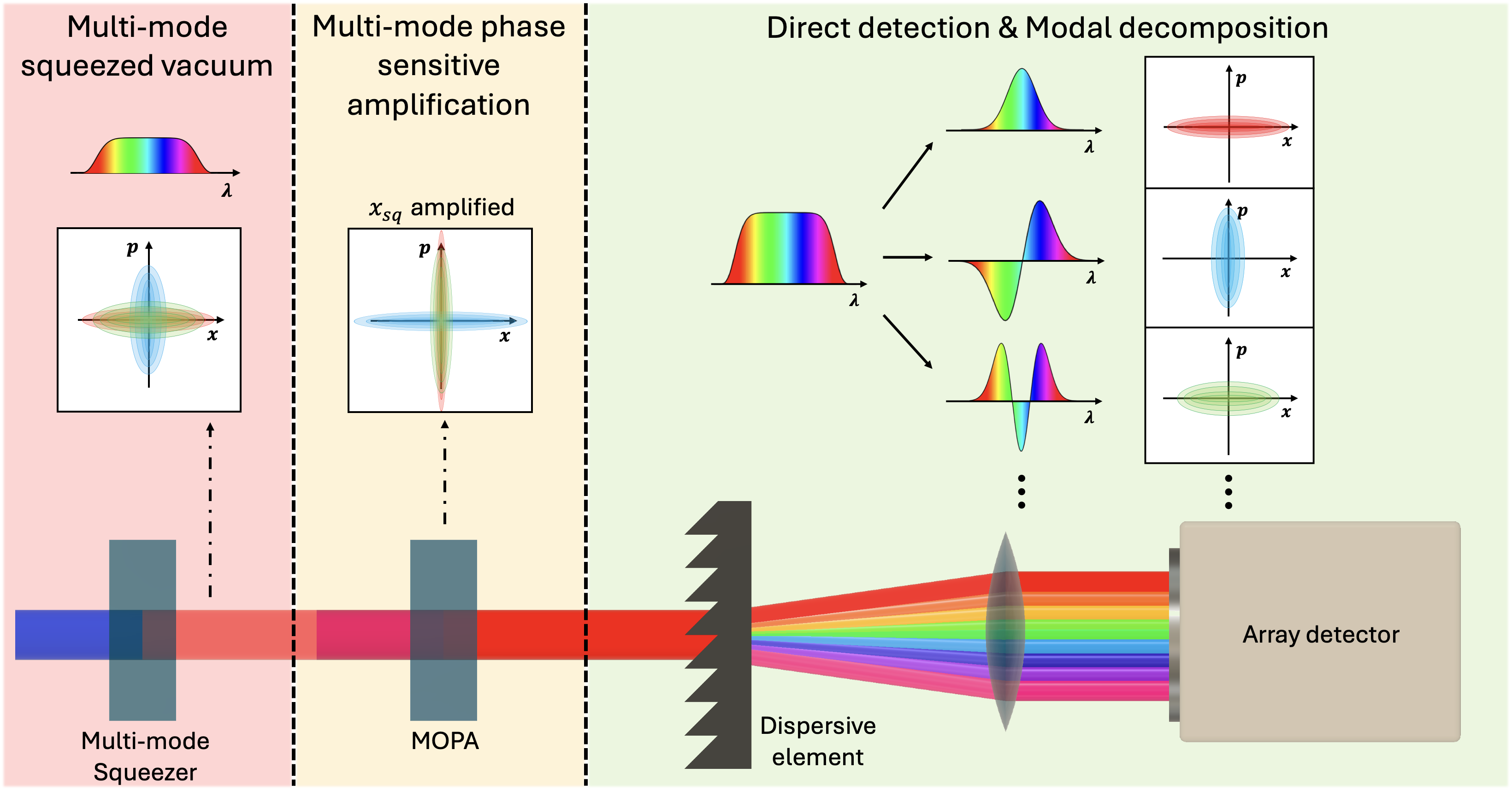}
     \caption{Simplified experimental scheme. A spectrally multimode squeezed vacuum is generated via type-I collinear-degenerate parametric down-conversion. Each mode undergoes quadrature amplification depending on the phase of the MOPA pump. A dispersive element (grating) then spatially separates the spectral components of the amplified radiation. Finally, spectral intensity correlations are measured and processed through a modal decomposition procedure, enabling the reconstruction of multimode squeezing information.} 
    \label{scheme}
\end{figure*}
Consider a multimode squeezed vacuum state at the input of MOPA, with photon annihilation operators $a_l$ acting in eigenmodes (also called Schmidt modes or squeezed modes) $u_l$ (Fig.\ref{idea}). Since the role of the MOPA is to retrieve information about the input fields, it is most convenient to describe the amplification process through the inverse Bogoliubov transformations~\cite{Barakat2025Feb,Scharwald2025}
\begin{equation}
    a_l=\sum_ng_{ln}[A_n \text{cosh}(G_n)-\text{e}^{-i\phi} A^{\dagger}_n\text{sinh}(G_n)].\label{amp}
\end{equation}
Here, $A_n$ is the annihilation operator associated with mode $v_n$ of the amplifier, $G_n$ is the amplification gain for mode $A_n$, and $g_{ln}$ marks the overlap between the input and the amplifier modes. For simplicity, we assumed $G_n$ to be high enough so that the modes of the phase-sensitive amplification (PSA) signal match those of the amplifier~\cite{Barakat2025Feb,Scharwald2025}. Equation (\ref{amp}) indicates that the information about an input mode $a_l$ is distributed across the various output modes $A_n$ of the amplifier (Fig. \ref{idea}). Consequently, the information lost in one output mode due to imperfect mode matching can be recovered from the other output modes. Therefore, for squeezing detection, it is sufficient to retrieve the  intensity contributions from all output modes to fully access the squeezing of the input modes. Indeed, the variance of the input quadrature $x_l^\phi$, associated with field $a_l$, is then inferred from the output phase-sensitive intensities $I_n^\phi(\lambda)$ as~\cite{Barakat2025Feb}
\begin{equation}
    \text{Var}(x_l^\phi) = \sum_n g_{ln}^2\text{e}^{-2G_n}I_n^\phi(\lambda).\label{sq}
\end{equation}
Importantly, here the $g_{ln}$ matrix is assumed to be real, which is only true if the phases acquired by different modes between the squeezer and the amplifier are the same~\cite{Barakat2025Feb}. For frequency modes, this condition is satisfied if the dispersion before MOPA is either negligible or compensated for. 
\begin{figure}[t!]
\centering
\includegraphics[width=0.91\linewidth]{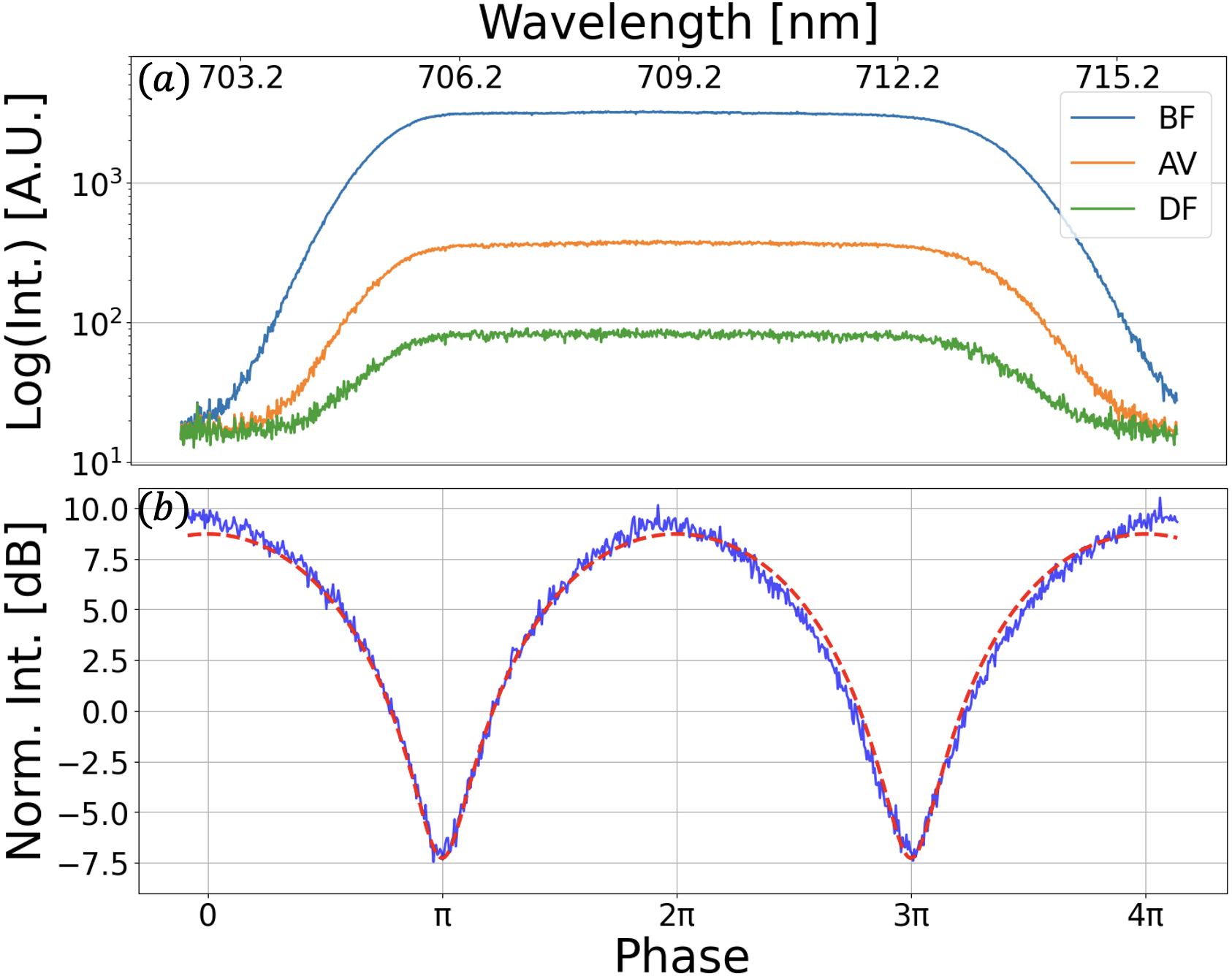}
     \caption{(a) Spectrum of bright fringe (blue)/dark fringe (green) compared to amplified vacuum. (b) Integral intensity, normalized to the one for amplified vacuum, as a function of the OPA pump phase (blue). The red dashed line shows a sinusoidal fit.} 
    \label{spectrum}
\end{figure}
To find the degree of squeezing of the input modes from Eq.~(\ref{sq}), one requires factors $\text{e}^{-2G_n}$ in its right-hand side. These factors are obtained by seeding the MOPA with the vacuum (i.e., blocking its input), measuring the vacuum quadratures $\text{Var}(x_l^{vac})$ in the same fashion, and normalizing $\text{Var}(x_l^\phi)$ to these values. 
\begin{figure*}[t!]
\centering
\includegraphics[width=0.9\linewidth]{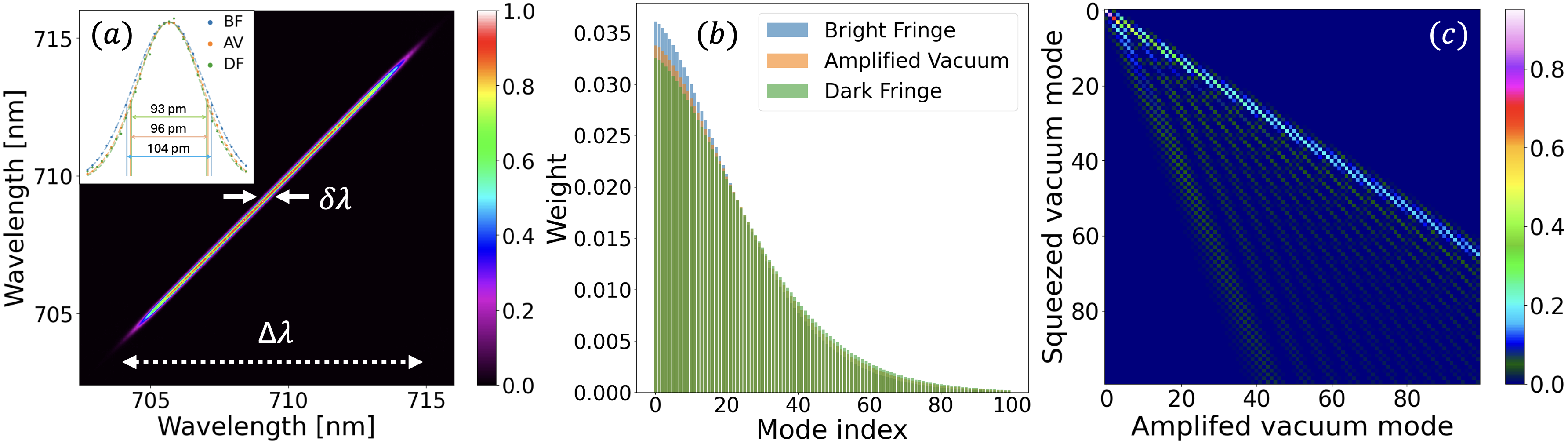}
     \caption{(a) Intensity covariance function for the amplified vacuum with $\Delta\lambda$ and $\delta\lambda$ denoting its unconditional and conditional widths, respectively. The inset shows the cross-sections (conditional distributions) of intensity covariance for the bright fringe (blue), amplified vacuum (orange), and dark fringe (green).  (b) Mode weights $w_n$ for BF, AV, and DF. (c) The mode overlap matrix $|g_{ln}|^2$ calculated numerically.} 
    \label{cov}
\end{figure*}

Crucially, this highlights that the detection of squeezing through MOPA does not require matching the amplifier modes to the input ones, as long as the output amplifier modes to which the input contributes are accessible.  This represents a significant advantage over the standard homodyne detection, where mode-matching between the local oscillator and the input mode is essential for a proper squeezing measurement. For highly multimode systems, homodyne detection becomes time-consuming and experimentally challenging. In contrast, MOPA enables simultaneous detection of all examined modes in a single measurement, making it far more efficient for high-dimensional applications.

\section{\label{exp}Experiment}

In the experiment (Fig. \ref{idea}), we generate spectrally multimode squeezing through collinear degenerate type-I (o->ee) parametric down-conversion (PDC) in a 3 mm Bismuth Triborate (BiBO) crystal. The pump is an 18 ps, 1 kHz laser at 354.67 nm, with a pulse energy of about 70 $\mu J$. The modal content of the squeezed state is determined by both the spectral profile of the pump and the phase-matching spectrum of the squeezer, and depends on the parametric gain~\cite{Scharwald2025,christ2013}. With the latter set to $G_{sq}=1.1\pm0.1$ for collinear emission, we generate > 500 squeezed spectral modes over a bandwidth of 37 nm, with the fundamental mode expected to exhibit a squeezing level of -10 dB. A typical characteristic of this well-known source is that odd and even modes exhibit squeezing in orthogonal quadratures: for instance, odd modes are p-squeezed while even modes are x-squeezed~\cite{Wasilewski2006}. 

The squeezed state is then fed into the MOPA, which is the same nonlinear crystal but pumped stronger to achieve a higher gain $G=4.5\pm0.2$. Given these parameters, the MOPA possesses effectively > 300 spectral modes. To achieve negligible dispersion, we filter only a small part of the squeezed light spectrum (8 nm around the degenerate frequency), which, however, contains more than 60 modes. The phase $\phi$ of the MOPA phase-sensitive amplification is introduced in the pump beam. The amplified radiation is then impinging on a single-mode fiber, filtering out only the fundamental Gaussian spatial  mode. Finally, a diffraction grating followed by an sCMOS camera allows access to the spectral components of the amplified radiation with a resolution of 13 pm. 

Fig.~\ref{spectrum}(a) shows the spectrum $I^\phi(\lambda)$ after the MOPA for phases $\phi$ at which MOPA amplifies the anti-squeezed quadrature (bright fringe (BF), blue), the squeezed quadrature (dark fringe (DF), green), and vacuum (orange). The latter is obtained by blocking the squeezed state at the input of the MOPA and thus shows the spectral and modal properties of the amplifier.  As shown, $I^\phi(\lambda)$ at the DF is consistently compared to that of the amplified vacuum across the entire spectral range. This indicates squeezing over the whole addressed spectral range, demonstrating the broadband character of both the squeezer and the MOPA. Further confirmation is provided by the PSA traces in Fig. \ref{spectrum}(b), where the normalized integrated intensity over the addressed spectrum reveals squeezing levels ranging from -7.2 dB to 9.5 dB. The red dashed line shows a sinusoidal fit. 
\section{spectrally multimode Squeezing}
Retrieving the multimode squeezing requires access to the intensities of the individual amplified modes (Eq. \ref{sq}). While one approach is spectral mode sorting, this was not available in our experiment. Instead, we applied the coherent mode decomposition directly to the amplified radiation in all three cases, amplified vacuum, BF, and DF, enabling the extraction of modal information through post-processing. To find how the output intensity is distributed over different modes, we use the fact that the Schmidt modes of squeezed vacuum, which combines signal and idler subsystems, are the same as the coherent modes of each subsystem taken separately~\cite{finger2017,Frascella2019Oct,Averchenko2020Nov}. Meanwhile, both signal and idler subsystems of squeezed vacuum are in a thermal state, and their coherent mode decomposition can be achieved through the measurement of the spectral intensity covariance function,
\begin{equation}
\text{Cov}(\lambda,\lambda')=\langle I(\lambda)I(\lambda')\rangle-\langle I(\lambda)\rangle\langle I(\lambda')\rangle.
\end{equation}

To select just one part of the signal–idler bipartite system, we slightly detune the single-mode filtered by the fiber from the one collinear with the pump. in experiment, we just coupled half of the full beam. 
\begin{figure*}[t!]
\centering
\includegraphics[width=0.9\linewidth]{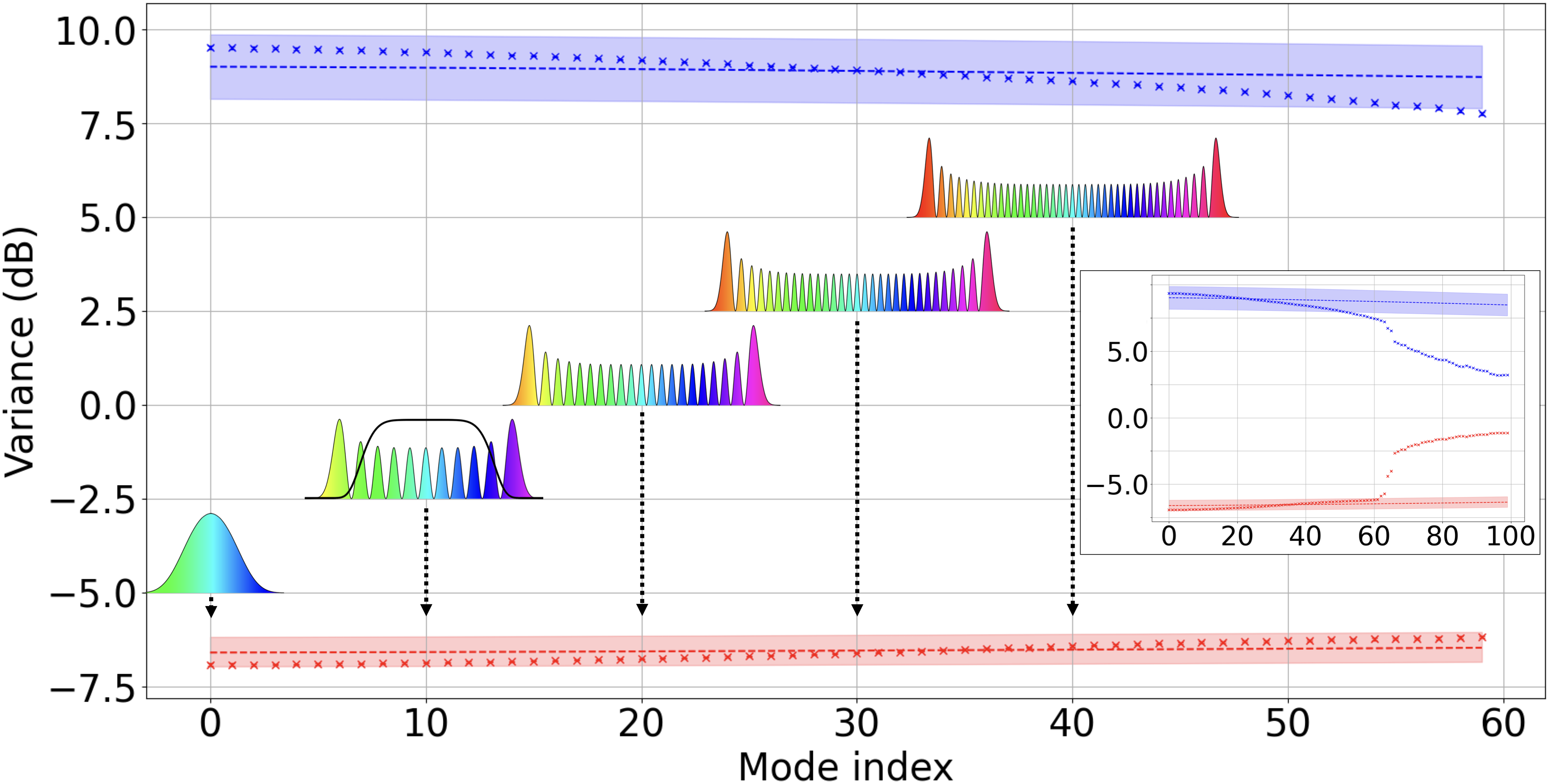}
     \caption{Simultaneously reconstructed squeezing/anti-squeezing levels (crosses) of each mode (up to 60; inset: up to 100). Dashed line shows the theoretical values, with the shaded area taking into account the uncertainty in $G_{sq}$ value. Spectral intensity distributions are shown for the fundamental, 10th, 20th, 30th, and 40th modes. A black line over the 10th mode highlights the spectral range used for filtering.}  
    \label{squeezingmodes}
\end{figure*}
Fig. 4(a) shows the measured spectral covariance function for amplified vacuum (AV). The large aspect ratio of the distribution indicates the highly multimodal character of the MOPA. The unconditional width $\Delta\lambda$ is mainly given by the spectral width of the bandpass filter. Therefore, in the DF and BF cases it is the same as in the case of AV.   
On the other hand, the conditional width $\delta\lambda$, indicating the correlations of spectral components, is almost the same for the  DF ($\delta\lambda=93 \ pm$) and AV ($\delta\lambda=96 \ pm$) but noticeably broader for the BF ($\delta\lambda=104 \ pm$), as shown in the inset of Fig. \ref{cov}(a). 
From the three intensity covariance functions, we infer the modal content of the amplified radiation as 
\begin{equation}
Cov(\lambda,\lambda')=|\sum_{m} w_{m} u_{m}(\lambda) u^{*}_{m}(\lambda')|^2,
\end{equation}
with $u_m(\lambda)$ and $w_m$ being the coherent modes and their corresponding weights respectively. These weights, shown in Fig. \ref{cov}(b) for the three different cases, provide the modal intensities $I_n^\phi(\lambda)=w_n^\phi I^\phi(\lambda)$ contributing to the amplified spectrum $I^\phi(\lambda)$. 
To retrieve squeezing from Eq.~(\ref{sq}), we need the overlap matrix $g_{ln}$ between the squeezed and amplifier modes. Figure \ref{cov}(c) shows $|g_{ln}|^2$ calculated for the $G_{sq}$ and $G$ values using integro-differential equations\cite{Barakat2025Feb,Scharwald2025}. Because $G_{sq}>G$, the modes of the squeezer are narrower, making them overlap with higher-order modes of the amplifier. 

Finally, from Eq.~(\ref{sq}), we find the degrees of squeezing for more than 60 spectral modes  (Fig. \ref{squeezingmodes}). Red (blue) crosses are experimentally obtained squeezing (anti-squeezing) values. Dashed lines are values calculated for the squeezer assuming $G_{sq}$ and 12\% effective losses. Shaded areas take into account the uncertainty in the measurement of $G_{sq}$. The inset shows the spectral intensity distributions $|u_n(\lambda)|^2$ of the 1st, 10th, 20th, 30th, and 40th modes. Since higher-order modes extend outside the filtered 8 nm bandwidth (see the 10th mode inset), the obtained squeezing and anti-squeezing values for these modes deviate from their theoretical values. Indeed, the low signal-to-noise ratio observed for modes beyond the 60th order causes the results to rapidly degrade (inset of Fig. \ref{squeezingmodes}). Nevertheless, the results show almost constant squeezing ranging between -7 dB and -6.5 dB up to the 60th mode, demonstrating the highly multimode nature of our squeezer and MOPA. This is of high interest for high-dimensional applications, especially cluster-based ones~\cite{pysher2011,chen2014,wang2025}. 
\\Interestingly, each of the cited experiments~\cite{pysher2011,chen2014,wang2025}, though distinct, produced 60-mode cluster states—making our current MOPA configuration a perfect candidate for addressing such states.


\section{Conclusion}
In conclusion, we have performed simultaneous detection of squeezing in 60 modes of spectrally multimode squeezed vacuum using multimode optical parametric amplification (MOPA) followed by direct detection. In our approach, MOPA transforms the variances of the input quadratures into intensity, enabling direct detection with a camera. Importantly, squeezing can be retrieved even when the modes of MOPA and squeezed vacuum are not matching, by taking into account their mode overlap.
Among hundreds of spectral modes, we observed a nearly constant squeezing level of 6.5 dB across all 60 modes. The number of measured modes was mainly limited by the use of an 8 nm bandpass filter, which was necessary to maintain a constant phase across the spectrum. We expect that dispersion compensation technique will enable us to observe squeezing over an even broader spectral range. 

Our work expands the scope of MOPA-based squeezing detection method from spatial to spectral modes, demonstrating its versatility regardless of the modal degree of freedom. It also highlights MOPA's ability to simultaneously measure a large number of modes, which offers a significant advantage for high-dimensional continuous-variable quantum information technologies.\newpage
\bibliography{Modes}

\end{document}